\documentclass[aps,prd,superscriptaddress,onecolumn,notitlepage,preprintnumbers]{revtex4-1}

\usepackage[english]{babel}
\usepackage{amsmath,graphicx,verbatim,epsfig}
\usepackage{epstopdf}
\usepackage{amsmath,graphicx}
\usepackage{color}
\usepackage{datetime}
\usepackage[colorlinks=true,linkcolor=blue,citecolor=blue]{hyperref}
	
\usepackage{subfig}
\usepackage{mathtools}

\usepackage{afterpage}

\newcommand{\sect}[1]{section~\protect\ref{#1}}

\newcommand{\clr}[1]{^{#1\hspace{-1px}}}

\newcommand{\vek}[1] {\boldsymbol{#1}}

\newcommand{\half}{{\textstyle\frac{1}{2}}}

\newcommand{\ms}{\mskip 1.5mu}

\allowdisplaybreaks[3]

\begin{document}

\preprint{NIKHEF 2014-041}


\title{Constraining double parton correlations and interferences}
  
\author{Tomas Kasemets}
\email{kasemets@nikhef.nl}
\affiliation{Nikhef and Department of Physics and Astronomy, VU
  University Amsterdam, De Boelelaan 1081, 1081 HV Amsterdam, The
  Netherlands}

\author{Piet J. Mulders}
\email{mulders@few.vu.nl}
\affiliation{Nikhef and Department of Physics and Astronomy, VU
  University Amsterdam, De Boelelaan 1081, 1081 HV Amsterdam, The
  Netherlands}



\begin{abstract}
Double parton scattering (DPS) has become very relevant as a background to interesting analyses performed by the experiments at the LHC. It encodes knowledge of correlations between the proton constituents not accessible in single parton scattering. Within perturbative QCD DPS is described in terms of partonic subprocesses and double parton distributions (DPDs). There exists a large number of different DPDs describing the different possible states of two partons inside a proton. They include correlations between the two partons and interferences between the two hard subprocesses. Taking the probability interpretation of the DPDs as starting point, we derive limits on the interference DPDs and thereby constrain the size of correlations between two partons inside an unpolarized proton.
\end{abstract}

\maketitle

\section{Introduction}
\label{sec:intro}

Hadron collisions with two hard subprocesses, double parton scattering (DPS), have become very relevant with the realization that they constitute an important background to many analyses at the LHC, such as Higgs boson coupling measurements and new physics searches involving high-multiplicity final states. The rapid growth in the density of partons with energy leads to a rapid increase of DPS cross sections, which intuitively increases as the parton density to the power of four. The description of DPS has seen important improvements over the last couple of years, moving towards a reliable treatment within perturbative QCD (see for example \cite{Bansal:2014paa,Diehl:2011tt,Bartalini:2011jp,Kasemets:2013nma}). Several ingredients in a proof of factorization have been established where the two hard subprocesses are calculated perturbatively, while the long distance physics is captured in double parton distributions (DPDs) \cite{Manohar:2012jr,Diehl:2011yj}.

Model estimates for the LHC based on the assumption of no correlations between the two hard subprocesses has been calculated for a variety of different processes. Double $c\bar{c}$ production and same sign double $W$-boson production are among the most promising for a clean separation of DPS from single parton scattering backgrounds \cite{Luszczak:2011zp,Gaunt:2014rua,Gaunt:2010pi,vanHameren:2014ava,Maciula:2013kd}. The fraction of DPS events at the LHC in the $W$-boson plus dijet final state has recently been measured by both ATLAS and CMS \cite{Chatrchyan:2013xxa,Aad:2013bjm}.

DPS has a rich structure and embodies features and challenges not present in single parton scattering. These arise from the correlations between the two partons inside a proton and the presence of interference between the two hard subprocesses. This includes interferences in the color, flavor, fermion-number and spin quantum numbers of the partons entering the two interactions. The origin of the interferences is simple: In single parton scattering the parton "leaving" a proton in the amplitude, has to have the same quantum numbers as the parton "returning" in the conjugate amplitude. In DPS it is only the sum of the quantum numbers in the amplitude which have to be matched in the conjugate amplitude.

These features are captured in a (perhaps dauntingly) large number of DPDs. The DPDs have been examined in a variety of different models, where correlations in general have been found to be sizable \cite{Rinaldi:2014ddl,Rinaldi:2013vpa,Broniowski:2013xba,Chang:2012nw}. The situation, however, is not quite as complex as it might seem at first sight, since in the canonical situations only a fraction of the DPDs are likely to play a significant role. Through detailed investigations an identification of and understanding for the relevant correlations in different processes and kinematic regions can be reached. DPS cross sections including quantum-number correlations have been calculated for double vector boson production (for any combination of $W$, $Z$ or $\gamma$) in the case of leptonic decay channels \cite{Kasemets:2012pr,Manohar:2012jr} and double $c\bar{c}$ production \cite{Kasemets:2014xxx}. Upper bounds on the polarized DPDs have been derived and through studies of their evolution the maximal degree of polarization of the two partons inside the proton at higher scales has been set \cite{Diehl:2013mla,Diehl:2014vaa}. The color correlations between two quarks within a proton are suppressed at large scales by evolution, with the physical interpretation being attributed to the transport of color over an hadronic distance inside the proton \cite{Manohar:2012jr,Mekhfi:1988kj}. This also affects the fermion-number interference between quarks and antiquarks or quarks and gluons. Flavor interference has so far been less studied.

In the present paper we derive upper bounds on the DPDs describing color, flavor and fermion-number interferences; thus constraining the correlations between two partons inside a proton, and their effect on cross sections. The bounds are based on the probability interpretation of the two parton densities, analogously to the Soffer bound \cite{Soffer:1994ww} for single parton distributions.

The structure of this paper is as follows: In \sect{sec:DPD} we review some of the basics of DPS relevant for our present purposes and give the definition of matrix elements for DPDs. In \sect{sec:cons} we derive constraints on the interference distributions in color, flavor and fermion-number for double parton distributions of quarks, antiquarks and gluons. We highlight the most important features and discuss our findings in \sect{sec:concl}.

\section{Double parton scattering and distributions}\label{sec:dps}
\label{sec:DPD}
The double parton scattering cross section can schematically be expressed as
\begin{align}
\frac{d\sigma}{\prod_{i=1}^2 dx_i d\bar{x}_i }\Bigg|_{DPS} & = \frac{1}{C} \hat{\sigma}_1\hat{\sigma}_2 \int d^2\vek{y}\ F(x_1,x_2,\vek{y}) \bar{F}(\bar{x}_1,\bar{x}_2,\vek{y})
\end{align}
where $\hat{\sigma}_i$ represents hard subprocess $i$, $C$ is a combinatorial factor equal to two (one) if the partonic subprocesses are (not) identical and $F$ ($\bar{F}$) labels the double parton distribution of the proton with momentum $p$ ($\bar{p}$). The DPDs depend on the longitudinal momentum fractions of the two partons $x_i$ ($\bar{x}_i$) and the distance between them $\vek{y}$. Implicit in this expression are the labels for the different flavors, colors, fermion numbers and spins of the four partons. This structure is significantly more complicated in DPS compared to the case with only one hard interaction, because of the possibility of interference between the two hard interactions and correlations between the two partons inside each proton.

The DPD for two partons in an unpolarized right-moving proton are defined as \cite{Diehl:2011yj}
\begin{align}
  \label{eq:dpds}
 F_{a_1a_2}(x_1,x_2,\vek{y})
 & = 2p^+ (x_1\ms p^+)^{-n_1}\, (x_2\ms p^+)^{-n_2}
        \int \frac{dz^-_1}{2\pi}\, \frac{dz^-_2}{2\pi}\, dy^-\;
           e^{i\ms ( x_1^{} z_1^- + x_2^{} z_2^-)\ms p^+}
\nonumber \\
 & \quad \times \left<p|\, \mathcal{O}_{a_2}(0,z_2)\, 
            \mathcal{O}_{a_1}(y,z_1) \,|p\right> \,,
\end{align}
where $n_i = 1$ if parton number $i$ is a gluon and $n_i = 0$ otherwise.
We use light-cone coordinates $v^\pm = (v^0 \pm v^3) /\sqrt{2}$ and the
transverse component $\vek{v} = (v\clr{1}, v^2)$ for any four-vector $v$.  
The operators for quarks read
\begin{align}
\label{eq:quark-ops}
  \mathcal{O}_{q_i}(y,z_i)
   &= \bar{q}_i\bigl( y - \half z_i \bigr)\,
       \Gamma_{q} \, q_i\bigl( y + \half z_i \bigr)
   \Big|_{z_i^+ = y^+_{\phantom{i}} = 0,\; \vek{z}_i^{} = \vek{0}},
\end{align}
with projection $\Gamma_q  = \half \gamma^+$ for unpolarized quarks.  The field with
argument $y + \half z_i$ in $\mathcal{O}_{q}(y,z_i)$ is associated with
a quark in the amplitude of a double scattering process and the field with
argument $y - \half z_i$ with a quark in the complex conjugate amplitude. The operators for an antiquark picks up an extra minus sign from interchanging the order of the fields.
The operators for gluons are
\begin{align}
\label{eq:gluon-ops}
  \mathcal{O}_{g_i}(y,z_i)
   &= \Pi_{g}^{ll'} \, G^{+l'}\bigl( y - \half z_i \bigr)\,
        G^{+l}\bigl( y + \half z_i \bigr)
   \Big|_{z_i^+ = y^+_{\phantom{i}} = 0,\; \vek{z}_i^{} = \vek{0}},
\end{align}
with projection $ \Pi_g^{ll'}  = \delta^{ll'}$ onto unpolarized gluons. 
We will take the two partons to be unpolarized throughout this paper, since the polarized distributions have already been studied in \cite{Diehl:2013mla}. We do not write out the Wilson lines that make the operators gauge invariant.

In analogy to the collinear single-parton distributions the DPDs can be interpreted as probability densities for finding two partons inside an unpolarized proton, with longitudinal momentum fractions $x_1$ and $x_2$ and transverse separation $\vek{y}$. As for single parton densities, this interpretation does not strictly hold in QCD where subtractions from the ultraviolet region can in principle invalidate the positivity, but it is nonetheless useful to investigate the consequences of the probability interpretation in order to guide the development of physically intuitive models of the distributions. This is particularly relevant in working at leading order of $\alpha_s$ where the connection between parton distributions and cross sections are most direct.

\section{Constraints on the interference distributions}
\label{sec:cons}
Since the probability density for finding two partons in a general
color or flavor state is positive semi-definite, we have
\begin{equation}
  \label{eq:pos-def-matrix}
\sum_{\lambda_1'\lambda_2'\ms \lambda_1^{}\lambda_2^{}}
 v^*_{\lambda_1'\lambda_2'} \,
 \rho_{(\lambda_1'\lambda_2')(\lambda_1^{}\lambda_2^{})}^{\phantom{*}} \,
 v_{\lambda_1^{}\lambda_2^{\phantom{\prime}}}^{\phantom{*}}
 \,\geq\, 0
\end{equation}
with arbitrary complex coefficients $v_{\lambda_1\lambda_2}$ normalized as
$\sum_{\lambda_1\lambda_2}|v_{\lambda_1\lambda_2}|^2 = 1$. $\lambda_i$ ($\lambda_i'$) labels the quantum numbers (colors or flavors) of the two partons in the (conjugate) amplitude. $\rho$ represent the color or flavor density matrix which are therefore positive-semidefinite. This property has already been
used for the spin density matrices associated with transverse-momentum
dependent distributions \cite{Bacchetta:1999kz}, generalized parton
distributions \cite{Diehl:2005jf} and double parton distributions \cite{Diehl:2013mla}. The positive semi-definiteness of the density matrix implies that the eigenvalues and principal minors are positive semi-definite, which leads to bounds on the elements of the matrix and thus on the DPDs. We will next go through the different types of interferences (i.e. color, flavor and fermion number) one by one and use the positivity to constrain the correlations between two partons inside an unpolarized proton.

\subsection{Quark color interference}
\begin{figure}[tb]
  \centering
  \includegraphics[width=0.4\textwidth]{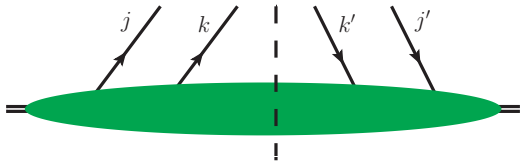}
\caption{Color labels of the double quark distributions. $j$ ($k$) labels the color of the quark taking part in the first (second) hard interaction in the amplitude. Primed indices refers to the conjugate amplitude.
}
\label{fig:col}
\end{figure}

The color structure of the double quark distributions in \eqref{eq:dpds} can be parametrized as \cite{Diehl:2011yj}
\begin{align}\label{eq:col}
F_{jj',kk'} = \frac{1}{N_c^2}\left[ \clr{1}F \; \delta_{jj'} \delta_{kk'} + \frac{2N_c}{\sqrt{N_c^2-1}}\,  \clr{8}F \;  t^a_{jj'}t^a_{kk'} \right],
\end{align}
where the unprimed indices $j,k$ corresponds to the quark in the amplitude entering the first and second hard interaction, while primed indices refers to the conjugate amplitude, as in figure~\ref{fig:col}. 

$\clr{1}F$ describe the distribution when the quark fields taking part in the same hard interaction form a color singlet, while $\clr{8}F$ describes the color octet interference contribution between the two hard interactions. The interference term couples the two quark fields in the same hard interaction in a color $SU(3)$ octet. The normalization of the interference distribution is chosen such that the two distributions enter with equal weight in the cross section for the production of a color singlet. With this choice the size of the two distributions directly indicates their phenomenological importance. For quark-antiquark (antiquark-quark) distributions the $k$ and $k'$ ($j$ and $j'$) indices in \eqref{eq:col} are interchanged. For DPDs with one quark and one antiquark the color structure is different and we will return to these distributions when discussing fermion-number interference in \sect{sec:ferm}.

Taking the three possible colors for the quarks ${r,g,b}$ and organizing the singlet and octet correlation distributions into a color-density matrix where the columns (rows) are given by the color of the two quarks in the (conjugate-) amplitude we could use the property of \eqref{eq:pos-def-matrix} to set limits on the color correlations. However, such a representation is reducible and we consider instead the two irreducible representations of two quarks in the product representation $3 \otimes 3 = \bar{3} \oplus 6$, i.e. in an anti-triplet and a sextet representation, diagonalizing the color density matrix.

Using a recoupling, the color anti-triplet and sextet distributions can be expressed in terms of the color square $\clr{1}F$ and correlation $\clr{8}F$ distributions by using projection operators \cite{Mekhfi:1985dv} $ (\delta_{jj'}\delta_{kk'} \pm \delta_{jk'}\delta_{kj'})$, where the minus (plus) sign gives the anti-triplet (sextet). The resulting relations reads
\begin{align}
F^{(\bar{3})}_{qq} & = \frac{1}{N_c^2} \left( \clr{1}F_{qq} - \frac{N_c+1}{\sqrt{N_c^2-1}}\, \clr{8}F_{qq} \right), &
F^{(6)}_{qq} & = \frac{1}{N_c^2} \left( \clr{1}F_{qq} + \frac{N_c-1}{\sqrt{N_c^2-1}}\, \clr{8}F_{qq} \right),
\end{align}
where we have chosen the normalization with respect to the $\clr{1}F_{qq}$ distribution. We will throughout this paper use the notation with a superscript following the distribution $F^{(R)}$ in order to denote the distribution in which the two partons in the amplitude couple to the $SU(3)$ representation $R$. This must not be confused with $\clr{R}F$, where the representation $R$ is formed by one field in the amplitude and one in the conjugate amplitude.
Up to overall normalization, equivalent equations can be obtained by making use of unitary $SU(3)$ color recouping matrices \cite{Jaffe:1976ih,Aerts:1979hn}.
The interpretation of $F_{qq}^{(\bar 3)}$ and $F_{qq}^{(6)}$ as describing probabilities for finding two quarks in definite color states gives the upper bounds on the color interference DPDs  
\begin{align}
 \clr{1}F_{qq} & \geq \frac{N_c+1}{\sqrt{N_c^2-1}} \,\clr{8}F_{qq}, &
  \clr{1}F_{qq} & \geq - \frac{N_c-1}{\sqrt{N_c^2-1}} \,\clr{8}F_{qq}.
\end{align}
These bounds apply also when both quarks are replaced by antiquarks. The bounds agree with the results found in \cite{Manohar:2012jr} when the difference in normalization of the interference distribution ($\clr{8}F$) is taken into account. 

\subsection{Flavor interference}
Specifying the flavor structure of the DPDs in equation \eqref{eq:dpds} gives, for example, the flavor squared distribution of an up and a down quark 
\begin{align}
 F_{ud}
  = 2p^+ 
        \int \frac{dz^-_1}{2\pi}\, \frac{dz^-_2}{2\pi}\, dy^-\;
           e^{i\ms ( x_1^{} z_1^- + x_2^{} z_2^-)\ms p^+}
 \left<p|\, (\bar{u}\; \Gamma_q\,  u) (\bar{d} \; \Gamma_q \, d) \,|p\right> \,,
\end{align}
while the corresponding flavor interference distribution is defined by
\begin{align}
 F^I_{ud}
 = 2p^+ 
        \int \frac{dz^-_1}{2\pi}\, \frac{dz^-_2}{2\pi}\, dy^-\;
           e^{i\ms ( x_1^{} z_1^- + x_2^{} z_2^-)\ms p^+}
\left<p|\, (\bar{u} \; \Gamma_q \, d) (\bar{d} \; \Gamma_q \, u) \,|p\right> \,.
\end{align}
Limiting ourselves to the first three quark flavors $u$, $d$ and $s$ we construct the flavor-density matrix 
\begin{align}
  \label{eq:col-matrix}
 \begin{pmatrix}
     F_{dd} & 
      0 & 
      0 & 
      0 & 
      0 & 
      0 & 
      0 & 
      0 & 
      0 
    \\[0.2cm]
    0 &
    F_{du} &
    0 &
    F^I_{ud} &
    0 &
    0 &
    0 &
    0 &
    0 
    \\[0.2cm]
    0 &
    0 &
    F_{ds} &
    0 &
    0 &
    0 &
     F^I_{sd} &
    0 &
    0 
        \\[0.2cm]
    0 &
    F^I_{du} &
    0 &
    F_{ud} &
    0 &
    0 &
    0 &
    0 &
    0 
        \\[0.2cm]
    0 &
    0 &
    0 &
    0 &
    F_{uu} &
    0 &
    0 &
    0 &
    0 
        \\[0.2cm]
    0 &
    0 &
    0 &
    0 &
    0 &
    F_{us} &
    0 &
    F^I_{su} &
    0 
        \\[0.2cm]
    0 &
    0 &
    F^I_{ds} &
    0 &
    0 &
    0 &
    F_{sd} &
    0 &
    0 
        \\[0.2cm]
    0 &
    0 &
    0 &
    0 &
    0 &
    F^I_{us} &
    0 &
    F_{su} &
    0 
        \\[0.2cm]
    0 &
    0 &
    0 &
    0 &
    0 &
    0 &
    0 &
    0 &
    F_{ss}
   \end{pmatrix},
\end{align}
where the columns (rows) correspond to the flavors of the two quarks in the (conjugate) amplitude, i.e. $dd$, $du$, $ds$, $ud$, $uu$, $us$, $sd$, $su$ and $ss$. Due to the positive semi-definiteness of the matrix \eqref{eq:pos-def-matrix} the principal minors of the two-dimensional subspaces gives
\begin{align}
F_{ab}F_{ba} \geq F^I_{ab}F^I_{ba},
\end{align}
while the positivity of the eigenvalues leads to the constraints on the flavor interference
\begin{align}
F_{ab}+F_{ba} \pm \sqrt{(F_{ab}-F_{ba})^2+4F^I_{ab}F^I_{ba}} \geq 0.
\end{align}

For mixed quark-antiquark distributions there can be flavor interference when the quark and antiquark are of the same flavor (i.e. $d\bar{d}$, $u\bar{u}$ or $s\bar{s}$). The flavor density matrix for quark-antiquark distributions reads
\begin{align}
  \label{eq:flav-matrix2}
 \begin{pmatrix}
     F_{d\bar{d}} & 
      0 & 
      0 & 
      0 & 
      F^{I_d}_{u\bar{u}} & 
      0 & 
      0 & 
      0 & 
      F^{I_d}_{s\bar{s}}  
    \\[0.2cm]
    0 &
    F_{d\bar{u}} &
    0 &
    0 &
    0 &
    0 &
    0 &
    0 &
    0 
    \\[0.2cm]
    0 &
    0 &
    F_{d\bar{s}} &
    0 &
    0 &
    0 &
    0 &
    0 &
    0 
        \\[0.2cm]
    0 &
    0 &
    0 &
    F_{u\bar{d}} &
    0 &
    0 &
    0 &
    0 &
    0 
        \\[0.2cm]
    F^{I_u}_{d\bar{d}} &
    0 &
    0 &
    0 &
    F_{u\bar{u}} &
    0 &
    0 &
    0 &
    F^{I_u}_{s\bar{s}} 
        \\[0.2cm]
    0 &
    0 &
    0 &
    0 &
    0 &
    F_{u\bar{s}} &
    0 &
    0 &
    0 
        \\[0.2cm]
    0 &
    0 &
    0 &
    0 &
    0 &
    0 &
    F_{s\bar{d}} &
    0 &
    0 
        \\[0.2cm]
    0 &
    0 &
    0 &
    0 &
    0 &
    0 &
    0 &
    F_{s\bar{u}} &
    0 
        \\[0.2cm]
    F^{I_s}_{d\bar{d}} &
    0 &
    0 &
    0 &
    F^{I_s}_{u\bar{u}} &
    0 &
    0 &
    0 &
    F_{s\bar{s}}
   \end{pmatrix},
\end{align}
where $F_{a\bar{a}}^{I_b}$ labels the distribution with $a\bar{a}$ in the amplitude and $b\bar{b}$ in the conjugate amplitude.
The principal minors of the two dimensional subspaces give the bounds on the interference distributions,
\begin{align}
F_{a\bar{a}}F_{b\bar{b}} \geq F_{a\bar{a}}^{I_b}F_{b\bar{b}}^{I_{a}}.
\end{align}
The most stringent bounds are given by the eigenvalues, which for the distributions where the quark and antiquark are of different flavors are simply the distributions (on the diagonal in the density matrix) themselves (there are no flavor interference for these distributions). For the distributions with quark-antiquark of equal flavor the eigenvalues result in rather complicated expressions, and there is little gain in giving them explicitly.

The bounds derived in this section are the first indication on the allowed size of the flavor interference DPDs and therewith their effect in DPS cross sections.

\subsection{Fermion-number interference}\label{sec:ferm}
For distributions with one quark and one anti-quark, there can further be interferences in fermion-number, i.e. between quarks and anti-quarks. These are always accompanied by color interference - since two quarks cannot couple to a color singlet. Therefore, in order to set limits on DPDs with a quark and an antiquark we consider the joint space of color and fermion-number. 

The DPDs describing interference in fermion-number between quarks and antiquarks are defined by \cite{Diehl:2011yj}
\begin{align}
	I_{q_1\bar{q}_2} (x_1,x_2,\vek{y})   & = 2p^+ 
        \int \frac{dz^-_1}{2\pi}\, \frac{dz^-_2}{2\pi}\, dy^-\;
           e^{i\ms ( x_1^{} z_1^- + x_2^{} z_2^-)\ms p^+}
\nonumber \\
 & \quad \times 
 \left<p|\, \bar{q}_2\bigl( \half z_2 \bigr)\,
     \half \gamma^+ \, q_1\bigl( y - \half z_1 \bigr)
     \bar{q}_2\bigl( -\half z_2 \bigr)\,
     \half \gamma^+ \, q_1\bigl( y + \half z_1 \bigr)  \,|p\right> \,,
\end{align}
and the color structure can be decomposed as
\begin{align}
I_{jj',kk'} = \frac{1}{N_c^2} \left( \clr{1}I \delta_{jk'}\delta_{j'k} + \frac{2N_c}{\sqrt{N_c^2-1}}\,  \clr{8}I t^a_{jk'}t^a_{j'k} \right).
\end{align}
As in the quark-quark case above we start by going to an irreducible representation, this time in terms of a singlet and an octet distributions $3 \otimes \bar{3} = 1 \oplus 8$. The quark-antiquark color density matrix reads
\begin{align}
\begin{pmatrix}
F^{(1)}_{q\bar{q}} & 0 \\[0.2cm]
0 & F^{(8)}_{q\bar{q}}
\end{pmatrix}.
\end{align}
The color singlet and octet distributions can be expressed in terms of $\clr{1}F$ and the interference distribution $\clr{8}F$ as
\begin{align}\label{eq:qqbar-rel}
F^{(1)}_{q\bar{q}} & = \frac{1}{N_c^2} \left( \clr{1}F_{q\bar{q}} + \sqrt{N_c^2-1} \, \clr{8}F_{q\bar{q}} \right) \geq 0, 
&
F^{(8)}_{q\bar{q}} & = \frac{1}{N_c^2} \left( \clr{1}F_{q\bar{q}} - \frac{1}{\sqrt{N_c^2-1}} \, \clr{8}F_{q\bar{q}}  \right) \geq 0
\end{align}
and their positivity leads to upper limits on the color correlations between the quark and antiquark. The inequalities hold true when interchanging $q$ and $\bar{q}$. The relations between $I^{(1)}$, $I^{(8)}$ and $\clr{1}I$, $\clr{8}I$ are equal to \eqref{eq:qqbar-rel} with $F$'s replaced by $I$'s, but as they are interference distributions they do not have to be positive. Considering next the joint space of color and fermion number, we can extend the two $2\times2$ matrices into a joint color-flavor number density matrix
where the columns (rows) correspond to the color representations and fermion numbers of the first and second parton in the (conjugate) amplitude, i.e. $q\bar{q}$ singlet, $q\bar{q}$ octet, $\bar{q}q$ singlet, $\bar{q}q$ octet,
\begin{align}
\begin{pmatrix}
F^{(1)}_{q\bar{q}} & 0 & I^{(1)}_{\bar{q}q} & 0 \\[0.2cm]
0 & F^{(8)}_{q\bar{q}} & 0 & I^{(8)}_{\bar{q}q} \\[0.2cm]
I^{(1)}_{q\bar{q}} & 0 & F^{(1)}_{\bar{q}q} & 0 \\[0.2cm]
0 & I^{(8)}_{q\bar{q}} & 0 & F^{(8)}_{\bar{q}q} 
\end{pmatrix}.
\end{align}
The eigenvalues of the matrix leads to the bounds
\begin{align}\label{eq:colnum}
F^{(1)}_{q\bar{q}} + F^{(1)}_{\bar{q}q} \pm \sqrt{(F^{(1)}_{q\bar{q}} - F^{(1)}_{\bar{q}q})^2 + 4 I^{(1)}_{q\bar{q}}I^{(1)}_{\bar{q}q}} &\geq 0,
&
F^{(8)}_{q\bar{q}} + F^{(8)}_{\bar{q}q} \pm \sqrt{(F^{(8)}_{q\bar{q}} - F^{(8)}_{\bar{q}q})^2 + 4 I^{(8)}_{q\bar{q}}I^{(8)}_{\bar{q}q}} &\geq 0.
\end{align}
 The combination of \eqref{eq:qqbar-rel} and its analogue for fermion-number interference distributions with the positive semi-definiteness of the eigenvalues leads to combined constraints on the fermion-number and color interference distributions of a quark and an anti-quark,
\begin{align}
	\clr{1}F_{\bar{q}q} & +  \clr{1}F_{q\bar{q}} - \frac{\clr{8}F_{\bar{q}q} +\, \clr{8}F_{q\bar{q}}}{\sqrt{N_c^2-1}} 
		\nonumber\\
		&\pm \sqrt{ \left( \clr{1}F_{\bar{q}q} -\, \clr{1}F_{q\bar{q}} +\frac{\clr{8}F_{q\bar{q}} - \, \clr{8}F_{\bar{q}q} }{\sqrt{N_c^2-1}} \right)^2 
			+ 4 \left(\clr{1}I_{\bar{q}q} - \frac{\clr{8}I_{\bar{q}q}}{\sqrt{N_c^2-1}}\right)\left(\clr{1}I_{q\bar{q}} - \frac{\clr{8}I_{q\bar{q}}}{\sqrt{N_c^2-1}}\right) }  \geq 0,
	\nonumber\\\nonumber\\
	\clr{1}F_{\bar{q}q} & + \clr{1}F_{q\bar{q}} + \sqrt{N_c^2-1} (\clr{8}F_{\bar{q}q} +\, \clr{8}F_{q\bar{q}}) 
	\nonumber\\
		& \pm \sqrt{  \left(\clr{1}F_{\bar{q}q}-\,\clr{1}F_{q\bar{q}}+\sqrt{N_c^2-1} (\clr{8}F_{\bar{q}q}-\,\clr{8}F_{q\bar{q}})\right)^2 
		  + 4\left(\clr{1}I_{\bar{q}q}+\sqrt{N_c^2-1} \,\clr{8}I_{\bar{q}q}\right)\left(\clr{1}I_{q\bar{q}}+\sqrt{N_c^2-1} \,\clr{8}I_{q\bar{q}}\right)}\geq 0.
\end{align}
The inequalities hold true when changing the $q\rightarrow \bar{q}$ and thus concludes our discussion of the DPDs with quarks and anti-quarks. We will now turn our attention to double gluon and mixed gluon-quark distributions.
%
\subsection{Double gluon distributions}
The double gluon distributions has a more elaborate color structure due to the increased number of representations when combining two color octets. Therefore, while discussing gluon distributions we for simplicity specify to $N_c=3$. Two color octets can then be combined into $8\otimes 8 = 1 \oplus 8_s \oplus 8_a \oplus 10 \oplus \overline{10} \oplus 27$, and we decompose the double gluon DPD as
\begin{align}\label{eq:glucol-dec}
F^{aa',bb'}  = \frac{1}{64} \Bigg[& \clr{1}F \delta^{aa'}\delta^{bb'} -\frac{\sqrt{8}}{3} \, \clr{A}F f^{aa'c}f^{bb'c} + \frac{3\sqrt{8}}{5} \, \clr{S}F d^{aa'c}d^{bb'c} + \frac{1}{\sqrt{5}} \, \clr{(10+\overline{10})}F \, (t_{10}^{aa',bb'} +t_{\overline{10}}^{aa',bb'})
	\nonumber\\
	&   + \frac{4}{\sqrt{27}} \, \clr{27}F t_{27}^{aa',bb'}\Bigg].
\end{align}
The projections onto the (anti-)decouplet and 27-tuple reads
\begin{align}
t_{10/\overline{10}}^{aa',bb'} &  = \delta^{ab}\delta^{a'b'} - \delta^{ab'}\delta^{a'b} -\frac{2}{3} f^{aa'c}f^{bb'c} \mp i(d^{abc}f^{a'b'c} +f^{abc}d^{a'b'c}), \nonumber\\
t_{27}^{aa',bb'} & = \delta^{ab}\delta^{a'b'} + \delta^{ab'}\delta^{a'b} -\frac{1}{4} \delta^{aa'}\delta^{bb'} - \frac{6}{5} d^{aa'c}d^{bb'c},
\end{align}
where the minus (plus) sign gives the $10$ ($\overline{10}$). The distributions $\clr{1}F$, $\clr{A}F$, $\clr{S}F$, $\clr{(10+\overline{10})}F$ and $\clr{27}F$ describe the case when the two gluon fields in the DPD which participate in the same hard interaction are coupled to a color singlet, an anti-symmetric octet, a symmetric octet, a decouplet or anti-decouplet and a 27-tuple. The decomposition in  \eqref{eq:glucol-dec} is different from what was done in \cite{Diehl:2011yj} in that it combines the decouplet and anti-decouplet into one distribution. The decouplet and anti-decouplet distributions are equal. As we explicitly show in \sect{app:gludist}, the equality $F^{(\overline{10})}_{gg}=F_{gg}^{(10)}$ is demonstrated by decomposing $F_{jj',kk.}$ in terms of $\clr{R}F$ (t-channel) distributions as in \eqref{eq:glucol-dec} and projecting onto the $F^{(R)}$ (s-channel) distributions. Reversing the role, decomposing in s-channel and projecting out the t-channel leads to $\clr{\overline{10}}F_{gg}=\,\clr{10}F_{gg}$ and the number of independent DPDs in \eqref{eq:glucol-dec} is reduced by one. The combination of decouplet and anti-decouplet has been discussed when projecting out color states of gluons in several other contexts, see for example \cite{Keppeler:2012ih,Kidonakis:1998nf,MacFarlane:1968vc,Bartels:1993ih,Oderda:1999kr}.

Coupling the color between the two gluons in the amplitude (and in the conjugate amplitude), using the projection operators in \cite{Mekhfi:1985dv}, gives
\begin{align}
	F_{gg}^{(1)} & =\frac{1}{64} \left[ \, \clr{1}F_{gg} + 2\sqrt{2} (\,\clr{S}F_{gg} - \,\clr{A}F_{gg}) + 2\sqrt{5}\,\clr{(10+\overline{10})}F_{gg} + 3\sqrt{3}\;\clr{27}F_{gg}\right] \geq 0,
	\nonumber\\
	F_{gg}^{(8a)} & = \frac{1}{64} \left[ \, \clr{1}F_{gg} + \sqrt{2} (\,\clr{S}F_{gg} - \,\clr{A}F_{gg}) - \sqrt{3}\,\clr{27}F_{gg} )\right] \geq 0,
	\nonumber\\
	F_{gg}^{(8s)} & = \frac{1}{64}\left[ \, \clr{1}F_{gg} - \frac{\sqrt{2}}{5} (3\,\clr{S}F_{gg} + 5\,\clr{A}F_{gg}) - \frac{4}{\sqrt{5}}\,\clr{(10+\overline{10})}F_{gg} + \frac{3\sqrt{3}}{5}\,\clr{27}F_{gg}\right] \geq 0,
	\nonumber\\
	F_{gg}^{(10+\overline{10})} & = \frac{1}{64} \left[ \, \clr{1}F_{gg} -\frac{4\sqrt{2}}{5}\,\clr{S}F_{gg} + \frac{2}{\sqrt{5}} \, \clr{(10+\overline{10})}F_{gg} -\frac{\sqrt{3}}{5} \,\clr{27}F_{gg} \right] \geq 0,
	\nonumber\\
	F_{gg}^{(27)} & = \frac{1}{64} \left[ \, \clr{1}F_{gg} + 2\sqrt{2}(\frac{1}{3}\,\clr{A}F_{gg} + \frac{1}{5}\,\clr{S}F_{gg})-\frac{2}{3\sqrt{5}} \,\clr{(10+\overline{10})}F_{gg} + \frac{7}{15\sqrt{3}} \,\clr{27}F_{gg} \right] \geq 0,
\end{align}
which describe the probability of finding two gluons inside a proton in a definite color state (color singlet, symmetric or anti-symmetric octet etc.) and their positivity leads to upper bounds on the color interference double gluon distributions. The projections were performed using the ColorMath package \cite{Sjodahl:2012nk} and a useful discussion on color projection operators can be found in \cite{Keppeler:2012ih}. 
\subsection{Mixed gluon quark distributions}
For the mixed gluon quark distributions the color decomposition reads \cite{Diehl:2011yj}
\begin{align}\label{eq:col-mixed}
F_{jj'}^{aa'} = \frac{1}{N_c(N_c^2-1)} \Bigg[  \,\clr{1}F \delta^{aa'}\delta_{jj'} - \,\clr{A}F \sqrt{2}i f^{aa'c}t^{c}_{jj'} + \sqrt{\frac{2N_c^2}{N_c^2-4}} \; \clr{S}F d^{aa'c}t^c_{jj'}  \Bigg].
\end{align}
Coupling the quark and the gluon in the (conjugate) amplitude we get the distributions of a gluon and a quark in the $8\otimes 3 = 3 \oplus 6 \oplus 15$ representations
\begin{align}\label{eq:gq-c2}
	F_{gq}^{(3)} & = \frac{1}{24} \left[ \,\clr{1}F_{gq} + \frac{1}{\sqrt{2}} (\sqrt{5}\,\clr{S}F_{gq} - 3\,\clr{A}F_{gq}) \right] \geq 0,
	\nonumber\\
		F_{gq}^{(6)} & = \frac{1}{24} \left[ \,\clr{1}F_{gq} - \frac{1}{\sqrt{2}} (\sqrt{5}\,\clr{S}F_{gq} + \,\clr{A}F_{gq}) \right] \geq 0,
	\nonumber\\
		F_{gq}^{(15)} & = \frac{1}{24} \left[ \,\clr{1}F_{gq} + \frac{1}{\sqrt{10}} (\,\clr{S}F_{gq} + \sqrt{5} \,\clr{A}F_{gq}) \right] \geq 0,
\end{align}
describing a quark and a gluon inside the proton in a color triplet, sextet or 15-tuple. Their positivity constraints the color interference distributions, however, just as for quark-antiquark distributions we can also have fermion number interference - now between a quark and a gluon. The fermion number interference describe when the gluon and quark have momentum fraction $x_1$ and $x_2$ respectively in the amplitude are interchanged in the conjugate amplitude (i.e. in the conjugate amplitude the quark have momentum fraction $x_1$ and the gluon $x_2$). This gives interference distributions $I_{jj'}^{aa'}$ with color structure decomposed as in \eqref{eq:col-mixed} and result in expressions for 
$I_{gq}^{(R)}$ with $R = \{3,6,15\}$ as in \eqref{eq:gq-c2} with $F\rightarrow I$ (with the difference that they do not have to be positive). This gives us the mixed color, fermion-number density matrix where the columns (rows) correspond to the color and fermion-number states of the quark and gluon in the (conjugate) amplitude
\begin{align}
\begin{pmatrix}
F^{(3)}_{gq} 	&	 0 	& 	0 	& I^{(3)}_{qg} 	& 	0 	& 0 \\[0.2cm]
0 & F^{(6)}_{gq} & 0 & 0 & I^{(6)}_{qg} & 0 \\[0.2cm]
0 & 0 & F^{(15)}_{gq} & 0 & 0 & I^{(15)}_{qg} \\[0.2cm]
I^{(3)}_{gq} 	&	 0 	& 	0 	& F^{(3)}_{qg} 	& 	0 	& 0 \\[0.2cm]
0 & I^{(6)}_{gq} & 0 & 0 & F^{(6)}_{qg} & 0 \\[0.2cm]
0 & 0 & I^{(15)}_{gq} & 0 & 0 & F^{(15)}_{qg} \\[0.2cm]
\end{pmatrix},
\end{align}
with eigenvalues
\begin{align}
	F^{(3)}_{gq} + F^{(3)}_{qg} \pm \sqrt{(F^{(3)}_{gq}-F^{(3)}_{qg})^2 +4I^{(3)}_{gq}I^{(3)}_{qg}} & \geq 0,
	\nonumber\\
	F^{(6)}_{gq} + F^{(6)}_{qg} \pm \sqrt{(F^{(6)}_{gq}-F^{(6)}_{qg})^2 +4I^{(6)}_{gq}I^{(6)}_{qg}} & \geq 0,
	\nonumber\\
	F^{(15)}_{gq} + F^{(15)}_{qg} \pm \sqrt{(F^{(15)}_{gq}-F^{(15)}_{qg})^2 +4I^{(15)}_{gq}I^{(15)}_{qg}} & \geq 0.
\end{align}
Replacing the distributions according to \eqref{eq:gq-c2} and its analogue for the fermion-number interference we obtain the bounds on the interference distributions, limiting the strength of color and fermion-number correlations between the quark and gluon inside the proton
\begin{align}
\clr{1}F_{gq} + \, \clr{1}F_{qg} & +\frac{1}{\sqrt{2}} \left( \sqrt{5}( \clr{S}F_{gq} +\, \clr{S}F_{qg}) - 3(\clr{A}F_{gq}+\,\clr{A}F_{qg}) \right)
	\nonumber\\
	& \pm \Bigg\{\left[\clr{1}F_{gq} - \, \clr{1}F_{qg}  +\frac{1}{\sqrt{2}} \left( \sqrt{5}( \clr{S}F_{gq} -\, \clr{S}F_{qg}) - 3(\clr{A}F_{gq} -\, \clr{A}F_{qg}) \right)\right]^2 
	\nonumber\\
	& \quad + 4\left(\clr{1}I_{gq}+\frac{1}{\sqrt{2}} (\sqrt{5} \, \clr{S}I_{gq}-3\, \clr{A}I_{gq})\right) 
		\left(\clr{1}I_{qg}+\frac{1}{\sqrt{2}} (\sqrt{5} \, \clr{S}I_{qg}-3\, \clr{A}I_{qg})\right)\Bigg\}^{1/2}\geq 0,
	\nonumber\\
	\clr{1}F_{gq} + \, \clr{1}F_{qg} & - \frac{1}{\sqrt{2}} \left( \sqrt{5}( \clr{S}F_{gq} +\, \clr{S}F_{qg}) + (\clr{A}F_{gq}+\,\clr{A}F_{qg}) \right)
	\nonumber\\
	& \pm \Bigg\{\left[\clr{1}F_{gq} - \, \clr{1}F_{qg}  - \frac{1}{\sqrt{2}} \left( \sqrt{5}( \clr{S}F_{gq} -\, \clr{S}F_{qg}) + (\clr{A}F_{gq} -\, \clr{A}F_{qg}) \right)\right]^2 
	\nonumber\\
	& \quad + 4\left(\clr{1}I_{gq}-\frac{1}{\sqrt{2}} (\sqrt{5} \, \clr{S}I_{gq}+\, \clr{A}I_{gq})\right) 
		\left(\clr{1}I_{qg}-\frac{1}{\sqrt{2}} (\sqrt{5} \, \clr{S}I_{qg}+\, \clr{A}I_{qg})\right)\Bigg\}^{1/2}\geq 0,
	\nonumber\\
	\clr{1}F_{gq} + \, \clr{1}F_{qg} & +\frac{1}{\sqrt{10}} \left( ( \clr{S}F_{gq} +\, \clr{S}F_{qg}) + \sqrt{5}(\clr{A}F_{gq}+\,\clr{A}F_{qg}) \right)
	\nonumber\\
	& \pm \Bigg\{\left[\clr{1}F_{gq} - \, \clr{1}F_{qg}  +\frac{1}{\sqrt{10}} \left(( \clr{S}F_{gq} -\, \clr{S}F_{qg}) +\sqrt{5} (\clr{A}F_{gq} -\, \clr{A}F_{qg}) \right)\right]^2 
	\nonumber\\
	& \quad + 4\left(\clr{1}I_{gq}+\frac{1}{\sqrt{10}} ( \clr{S}I_{gq} + \sqrt{5} \, \clr{A}I_{gq})\right) 
		\left(\clr{1}I_{qg}+\frac{1}{\sqrt{10}} ( \clr{S}I_{qg} + \sqrt{5} \, \clr{A}I_{qg})\right)\Bigg\}^{1/2}\geq 0.
\end{align}
These constraints are applicable also when the quark is replaced by an antiquark and thus completes our list of interferences in DPS.
\section{Conclusions}
\label{sec:concl}
The interpretation of double parton distributions as probabilities for finding two partons in an unpolarized proton has been used to constrain the strength of correlations between the two partons and the effect of interferences in double parton scattering. Limits have been set on the size of the color, flavor and fermion-number interference DPDs involving quarks, anti-quarks and gluons. Combined with the bounds on the polarized DPDs \cite{Diehl:2013mla} they constrain all the quantum-number interference types in double parton scattering. The constraints are interesting in that they make explicit the interdependence of the different distributions, for example Eq.~\eqref{eq:colnum} shows that an increased fermion-number interference decreases the maximal size of the difference between the the distributions of quark-antiquark compared to  antiquark-quark (both in a color singlet or octet state).

The limits can be useful in constructing models for the DPDs and in examining the possible correlation effects in DPS cross sections. The large number of different DPDs makes it cumbersome to take all interferences into account for phenomenological calculations, and unfeasible to extract all of them experimentally. The bounds provides a starting point when one considers the observable effects of the correlations and tries to determine which correlations have to be taken into account in phenomenological studies. For a particular process, the combination of the bounds with the knowledge of the evolution of the DPDs, can already lead to a large reduction of the relevant correlations in the process and the number of DPDs which should be included in the cross section calculation.

The effect of evolution on the bounds should be further investigated. Color interference and fermion-number interference for quarks and anti-quarks are suppressed in evolution to higher scales \cite{Manohar:2012jr}. By combining the derived constraints with the evolution of these distributions, upper limits can be set on the scale at which color interference can be of experimental relevance. Analogous suppression by evolution is expected for color interference in the gluon and mixed quark-gluon sectors, but the exact expressions remains to be worked out. The evolution of the flavor interference distributions has been less studied, but since these distributions does not mix with the gluon distributions they are expected to become less prominent in the small $x_i$, large $Q$ region.

In deriving the bounds, we showed that the two distributions for finding two gluons in a decouplet or an anti-decouplet are equal. They can therefore be combined in the decomposition of the double gluon color structure \eqref{eq:glucol-dec}, reducing the number of independent double gluon distributions.

%
\section*{Acknowledgements}
We thank Markus Diehl and Wouter Waalewijn for valuable discussions and input on the manuscript, and Mathias Ritzmann for useful discussions in particular on color projection operators. We acknowledge
financial support from the European Community under the ``Ideas'' program
QWORK (contract 320389). Figures were made using JaxoDraw \cite{Binosi:2003yf}.

\appendix

\section{Equivalence of double gluon decouplet and anti-decouplet distributions}\label{app:gludist}
We can show that the gluon decouplet and anti-decouplet distributions are equal by decomposing the color structure of the double gluon distribution with separate $\clr{10}F$ and $\clr{\overline{10}}F$ as in \cite{Diehl:2011yj}
\begin{align}
F^{aa',bb'}  = \frac{1}{64} \Bigg[& ^1F \delta^{aa'}\delta^{bb'} -\frac{\sqrt{8}}{3} \, ^AF f^{aa'c}f^{bb'c} + \frac{3\sqrt{8}}{5} \, ^SF d^{aa'c}d^{bb'c} + \frac{2}{\sqrt{10}} \, ^{10}F t_{10}^{aa',bb'} 
	\\
	&  + \frac{2}{\sqrt{10}} \,^{\overline{10}}F (t_{10}^{aa',bb'})^* + \frac{4}{\sqrt{27}} \, ^{27}F t_{27}^{aa',bb'}\Bigg].
\end{align}
The projections onto the $F^{(10)}$ and $F^{(\overline{10})}$ distributions reads
\begin{align}
F^{(10)} = \frac{1}{40}t_{10}^{ab,a'b'} F^{aa',bb'} & = \frac{1}{64} \left[ \, \clr{1}F + \sqrt{\frac{2}{5}} \left( \, \clr{10}F + \, \clr{\overline{10}}F\right) - \frac{4\sqrt{2}}{5} \, 		\clr{S}F - \frac{\sqrt{3}}{5} \, \clr{27}F \right] 
	\nonumber\\
F^{(\overline{10})} = \frac{1}{40}t_{\overline{10}}^{ab,a'b'} F^{aa',bb'} & = \frac{1}{64} \left[ \, \clr{1}F + \sqrt{\frac{2}{5}} \left( \, \clr{10}F + \, \clr{\overline{10}}F\right) - \frac{4\sqrt{2}}{5} \, 	\clr{S}F - \frac{\sqrt{3}}{5} \, \clr{27}F \right],
\end{align}
which shows that $F^{(10)} = F^{(\overline{10})}$. We can likewise decompose the color structure of $F^{aa',bb'}$ in terms of $F^{(10)}$ and $F^{(\overline{10})}$ and use $t_{10/\overline{10}}^{aa',bb'}$ to project out $\clr{10}F$ and $\clr{\overline{10}}F$. The resulting expressions show that $\clr{10}F=\, \clr{\overline{10}}F$. The two distributions can therefore be combined into one $\clr{(10+\overline{10})}F$ as in \eqref{eq:glucol-dec}.

\bibliography{../LatexStuff/KasBib}

\end{document}